\documentclass[%
 reprint,
superscriptaddress,
 amsmath,amssymb,longbibliography,
 aps,
prb,
]{revtex4-1}
\usepackage{hyperref}
\usepackage{graphicx}
\usepackage{amsmath}
\usepackage{color}
\usepackage{physics}

\newcommand{\sruo}{{Sr}$_2${RuO}$_4$}

\newcommand{\io}{\tilde \mu}

\newcommand{\down}{\downarrow}
\newcommand{\up}{\uparrow}
\newcommand{\spin}{\sigma}

\newcommand{\rv}{{\bf r}}

\newcommand{\kv}{{\bf k}}
\newcommand{\qv}{{\bf q}}

 \usepackage{tabularx}
\usepackage{pdfpages} 
\usepackage{pgffor} 
\makeatletter
\AtBeginDocument{\let\LS@rot\@undefined}
\makeatother

\def\supplementfilename{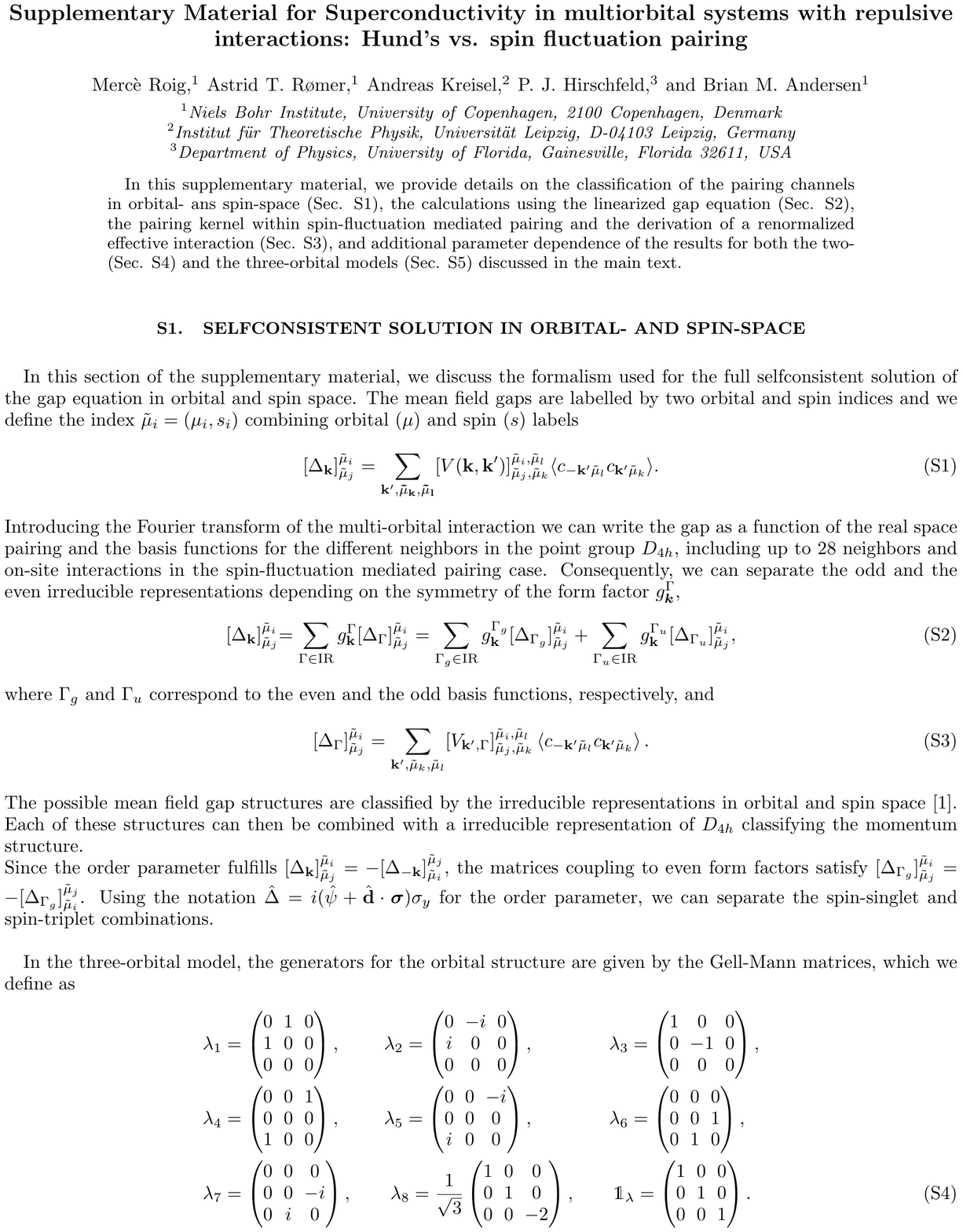}
\pdfximage{\supplementfilename}
\def\numbersupplementpages{\the\pdflastximagepages}

\newif\ifarXiv
\arXivtrue 
 
\begin{document}

\title{Superconductivity in multiorbital systems with repulsive interactions:\\ Hund's pairing vs. spin-fluctuation pairing}

\author{Merc\`e Roig}
\affiliation{Niels Bohr Institute, University of Copenhagen, 2100 Copenhagen, Denmark}

\author{Astrid T. R\o{}mer}
\affiliation{Niels Bohr Institute, University of Copenhagen, 2100 Copenhagen, Denmark}
 
\author{Andreas Kreisel}
 \affiliation{Institut f\"ur Theoretische Physik, Universit\"at Leipzig, D-04103 Leipzig, Germany}
 
\author{P. J. Hirschfeld}
\affiliation{Department of Physics, University of Florida, Gainesville, Florida 32611, USA}

\author{Brian M. Andersen}
\affiliation{Niels Bohr Institute, University of Copenhagen, 2100 Copenhagen, Denmark}


\begin{abstract}

Hund's pairing refers to Cooper pairing generated by onsite interactions that become attractive due to large Hund's exchange $J$. This is possible in multiorbital systems even when all local bare interactions are repulsive, since attraction in specific channels are given by certain linear combinations of interaction parameters. On the other hand,  pairing processes such as the exchange of spin fluctuations, are also present. We compare mean-field Hund's pairing and spin-fluctuation-mediated pairing using electronic bands appropriate for different classes of multiorbital systems over a wide range of interaction parameters. We find that for systems without clear nesting features, the superconducting state generated by the Hund's mechanism agrees well with that from the full fluctuation exchange vertex when Hund's exchange and spin-orbit coupling are sufficiently large.
On the other hand, for systems characterized by a peaked finite-momentum particle-hole susceptibility, spin-fluctuation pairing generally dominates over Hund's pairing. From this perspective Hund's pairing states appear unlikely to be realized in systems like Sr$_2$RuO$_4$ and generic iron-based superconductors.
\end{abstract}

\maketitle

\section{introduction}
 
The problem of creating superconductivity in systems where local interactions are repulsive dates back to Kohn and Luttinger, who studied how the screened Coulomb interaction can give rise to effective attraction in higher angular momentum channels in the electron gas\;\cite{Kohn1965}.  Electrons binding in such  pair states can avoid the Coulomb energy cost largely without the need of retardation in time, since the pair wave function has a node at the origin.  
In the electron gas, as in single-band lattice systems like the Hubbard model, the bare interaction
itself cannot pair directly.  Instead, one needs to calculate the effective interaction, i.e. sum screening processes leading to Friedel oscillations that allow for attraction in certain channels\;\cite{Kohn1965,berk,Fay,Scalapino1986,Scalapino1999,Miyake86,MaitiChubukov2013,RomerPRR2020}.

By contrast, many unconventional superconductors of current interest involve electrons in multiple bands at the Fermi level, subject to strong local Coulomb repulsion\;\cite{Scalapino_RMP}. This situation is described by a Hubbard-Kanamori Hamiltonian, with Hund's exchange $J$, inter-orbital Coulomb interaction $U'$, and pair hopping term $J'$ in addition to the usual intra-orbital Hubbard $U$.  In this model, the attraction binding electrons into Cooper pairs can arise in different ways. First, it can be generated as in the one-band case by summing the pair scattering processes.  In the iron-based superconductors, magnetic fluctuations connecting the Fermi surface pockets in the Brillouin zone (BZ) are thought to be exchanged, leading to dominant spin-singlet sign-changing $s_\pm$- or $d$-wave condensates\;\cite{Hirschfeld_2011,Chubukov2012,Kreisel_review}.

In recent years, a qualitatively different route to pair formation in multiorbital models with local repulsive interactions has been explored by a number of 
authors\;\cite{Spalek_2001,Han_2004,Zegrodnik_2014,Hoshino_2015,Kubo_2007,Puetter_2012,suh2019,clepkens_PRB_2021,Lee_2008,Xia_2008,Vafek17,Boker_2019,Cheung_2019,Sakai_2004}.
Projecting onto  appropriate symmetry channels, the interaction parameters $U,U',J,J'$ appear in certain combinations that may be negative and lead to direct attraction.
For example, for certain iron-based systems, an inter-orbital spin-triplet state is stabilized by an interaction channel proportional to $U'-J$, providing attraction when $J$ exceeds $U'$\;\cite{Vafek17} and spin-orbit coupling (SOC) is present to generate a Cooper log at the instability\;\cite{Hoshino_2015,Vafek17}. 
We refer to pairing of this type as ``Hund's pairing" since it requires a large Hund's exchange $J$. It has been suggested to produce exotic pair states in uranium-based superconductors\;\cite{Han_2004,Zegrodnik_2014,Hoshino_2015}, iron-based superconductors\;\cite{Xia_2008,Vafek17,Boker_2019} and Sr$_2$RuO$_4$\;\cite{Hoshino_2015,Kubo_2007,Puetter_2012,suh2019,clepkens_PRB_2021}.

Even if Hund's pairing states are stable, they still ``compete" with more usual spin-fluctuation-driven pairing states\;\cite{Romer3D}.
Until now, no one has compared the two types of pairing on an equal footing.
This is important, because although SOC creates a log instability, the corresponding $T_c$ may be very low because of the small weight of intra-band pairing present in the condensate.
One expects that for sufficiently large $J$ and SOC such states will indeed be stable, but can Hund's pairing out-compete spin-fluctuation driven pairing, or will the latter states remain favorable also at large $J/U$? And how might this depend on the underlying band structure? It is vital that the discussion regarding these candidate states in systems of current interest proceed with accurate estimates of their true viability.

In this work, we compare the superconducting order obtained from the two different pairing mechanisms: local Hund's pairing at the mean-field level vs. spin-fluctuation mediated pairing. The comparative study is performed for two very different multiorbital systems: 1) a two-orbital model forming two $\Gamma$-centered Fermi pockets\;\cite{Vafek17}, and 2) a three-orbital model relevant for Sr$_2$RuO$_4$\;\cite{Cobo16,RomerPRL}. In both procedures, we determine the gap structure and $T_c$ while exploring different interaction regimes, varying $U\mathrm{,}\;J/U$, and the value of SOC $\lambda_{so}$. As is well-known, Hund's pairing at the mean-field level is only operative in the regime, $J>U'$\;\cite{Puetter_2012,Vafek17}. This criterion can be renormalized by higher-order processes\;\cite{Hoshino_2015,Gingras18}, effectively expanding the ``Hund's regime". We find that inside this regime, pairing driven by Hund's coupling agrees almost quantitatively with spin-fluctuation pairing when the electronic structure generates a susceptibility with weak momentum structure. By contrast, when the susceptibility contains sufficient momentum structure, spin-fluctuation processes generally dominate the pairing, also in the Hund's regime. We analyze this limit in detail and explain why Hund's pairing and spin-fluctuation pairing can lead to both qualitatively and quantitatively different superconducting solutions. 

\section{Method}

Before applying specific band structures,
 we describe the methodology applied as pairing kernel for both mechanisms in question. The starting point is the multiorbital interaction 
\begin{align}
\hat H_{int}\!=\!\frac{1}{2}\!\!\sum_{ \kv,\kv' \{\tilde \mu\}}\!\!\Big[V(\kv,\kv')\Big]^{\io_i , \io_l }_{\io_j,\io_k } c_{\kv \io_i }^\dagger c_{-\kv \io_j }^\dagger c_{-\kv' \io_l } c_{\kv' \io_k },
\label{eq:Hcooper}
\end{align}
where $c_{\kv\io}$ annihilates an electron with momentum $\kv$, and $\io=(\mu,\spin)$ is a joint index of orbital and electronic spin.

We refer to Hund's pairing when onsite interactions directly mediate superconductivity. Thus, for Hund's pairing the effective interaction of Eq.\;(\ref{eq:Hcooper}) is given simply by
$[V(\kv,\kv')]^{\io_i , \io_l }_{\io_j,\io_k } =[U]^{\io_i , \io_l }_{\io_j,\io_k }$ which
contain intra- ($U$) and inter-orbital Coulomb scattering ($U'$) as well as pair-hopping terms ($J,J'$) with the spin rotational invariant setting $U^\prime=U-2J\mathrm{,}\;J^\prime=J$, for details see the supplementary material (SM)\;\cite{Supplementary}.
We require throughout that $U'\geq 0$, i.e. $J/U\leq\frac{1}{2}$. Attractive pairing emerges when $J>U'$ (equivalently $J/U>\frac{1}{3}$) at the mean-field level
which has been recently shown to generate unusual orbital-singlet, spin-triplet gap structures\;\cite{Puetter_2012,Hoshino_2015,Vafek17,clepkens_PRB_2021}.

Within spin-fluctuation mediated pairing, an effective interaction is derived from random phase approximation (RPA) diagrams\;\cite{Scalapino_RMP,Romer2015,romer2021}. The pairing obtained in this framework is given by 
\begin{eqnarray}
\Big[V(\kv,\kv')\Big]^{\io_i , \io_l }_{\io_j,\io_k } &\!\!=\!&\Big[U\Big]^{\io_i , \io_l }_{\io_j,\io_k }\!\!+\!\Big[U\frac{1}{1-\chi_0U}\chi_0U\Big]^{\io_i, \io_l}_{\io_j, \io_k}(\kv+\kv') \nonumber \\
&& -\Big[U\frac{1}{1-\chi_0U}\chi_0U  \Big]^{\io_i,\io_k}_{\io_j ,\io_l}(\kv-\kv') ,
\label{eq:Veff}
\end{eqnarray}
and includes, in addition to the same local interactions as the Hund's pairing scenario, explicit momentum-dependent effective interactions through the bare susceptibility $\chi_0(\qv)$, see SM for details\;\cite{Supplementary}. Below we compare the two mechanisms by solving (i) the resulting BCS gap equation for different bands and interaction parameters
\begin{equation}
    [\Delta_{\bf{k}}]^{\Tilde{\mu}_i}_{\Tilde{\mu}_j} = \sum_{\bf k',\Tilde{\mu}_k, \Tilde{\mu}_l} [V(\kv,\kv')]^{\Tilde{\mu}_i,\Tilde{\mu}_l}_{\Tilde{\mu}_j,\Tilde{\mu}_k} \langle c_{-\kv' \Tilde{\mu}_l} c_{\kv'\Tilde{\mu}_k} \rangle
\end{equation}
and (ii) analyzing the linearized gap equation (LGE) projected to band- and spin-space\;\cite{Romer2015,RomerPRL,romer2021,Supplementary} to visualize the gap function $\Delta_{l}(\kv_f)$ on the Fermi surface and discuss leading and sub-leading instabilities according to the eigenvalue $\lambda$.

\section{Results}

\subsection{Non-nested band structure}

\begin{figure}[tb]
    \includegraphics[width=\linewidth]{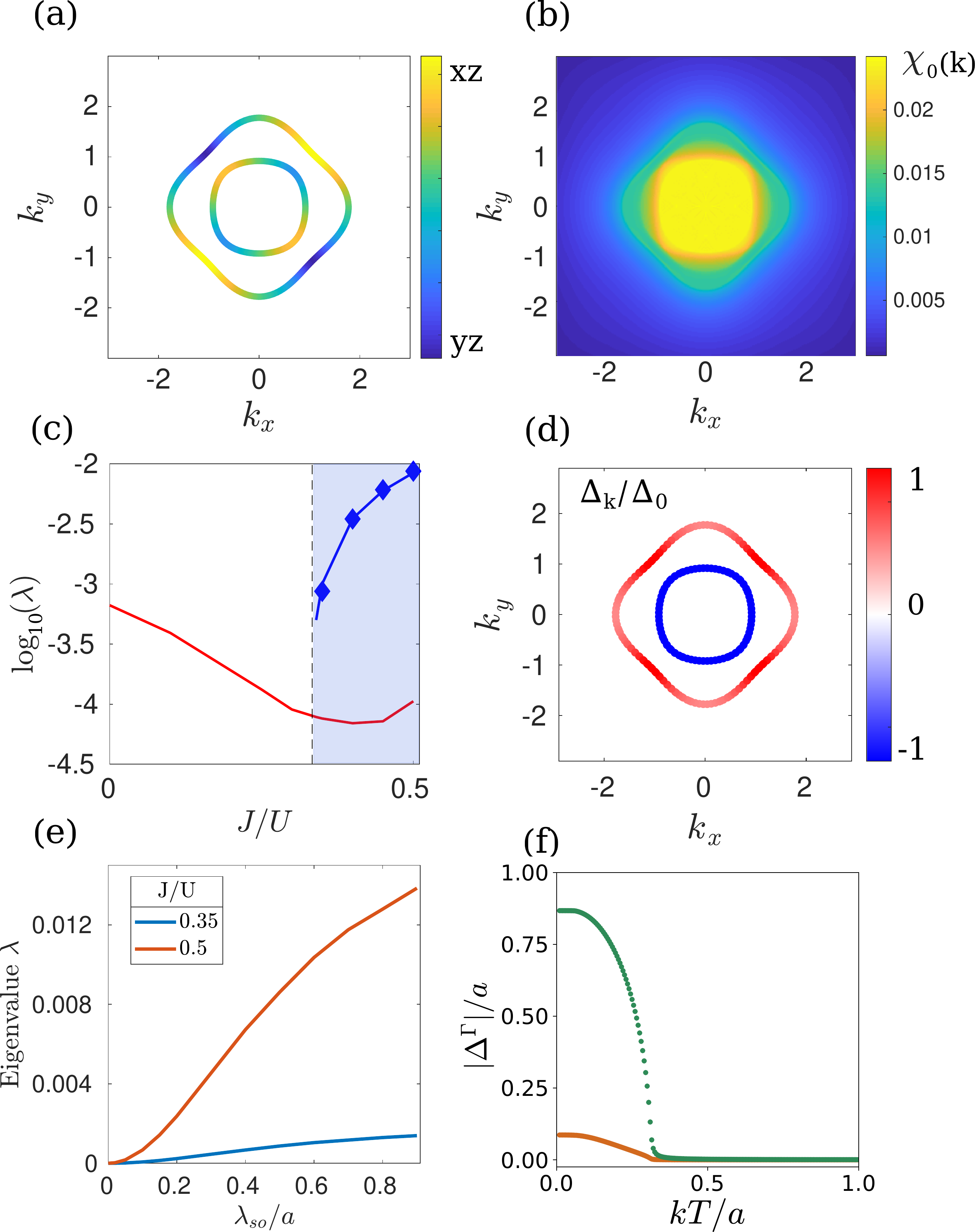}
    \caption{(a) Fermi surface of the two-orbital model Eq.\;(\ref{eq:H0_vafek}), with parameters: $a=1\mathrm{,}\;b=c=\lambda_{so}=0.5\mathrm{,}\;\mu=1.5$. The main orbital content is shown by the color code. (b) Momentum dependence of the static spin susceptibility $\chi_0^{zz}(\qv)$. (c) Leading eigenvalue of the LGE as a function of $J/U$ for $U/a=1$. Results from spin-fluctuation pairing (Hund's pairing) are indicated by solid lines (diamonds). Light blue region indicates regime of attractive onsite Hund's pairing, $J/U>\frac{1}{3}$. (d) Superconducting gap structure (in band space) obtained at $U/a=1\mathrm{,}\;J/U=0.5$, representative of all leading even-parity solutions when $J/U>\frac{1}{3}$. (e) Eigenvalue from the LGE as a function of $\lambda_{so}$ for different $J/U$ for Hund's pairing. (f) Orbital- and spin-structure (see Table\;\ref{tab:VafChu_selfcons_OP_channels}) from the Hund's pairing mechanism versus $T$ for $U/a=6\mathrm{,}\;J/U=0.5$, showcasing the dominant spin-triplet, orbital-singlet pairing channel (green curve) in the Hund's regime.}
    \label{fig:vafek}
\end{figure}

First, we discuss a multiorbital case with a simple band structure without nesting using the two-orbital model of Ref.\;\onlinecite{Vafek17}
\begin{align}
  H_0(\kv)\!=\!\!\left(\begin{array}{cc}
      \mu -a\kv^2+bk_xk_y & c(k_x^2-k_y^2)-i\sigma\lambda_{so} \\
      c(k_x^2-k_y^2)+i\sigma\lambda_{so} &  \mu -a\kv^2-bk_xk_y 
  \end{array}\right)
   \label{eq:H0_vafek}
   \end{align}
in the basis $[c_{\kv,xz,\sigma},c_{\kv,yz,\sigma}]$ with the energy unit $a=1/2m$.
The Fermi surface and the momentum structure of the bare static susceptibility $\chi_0^{zz}(\qv)$ is shown in Fig.\;\ref{fig:vafek}(a) and Fig.\;\ref{fig:vafek}(b), respectively.
At low values of $J/U<\frac{1}{3}$, the rather featureless susceptibility supports only weak (spin-fluctuation generated) superconductivity, as seen by the red curve in Fig.\;\ref{fig:vafek}(c). The favored nature of the superconductivity in this regime is helical odd-parity pairing. By contrast, in the Hund's regime where $J/U>\frac{1}{3}$, the regime of main interest in this paper, Hund's pairing becomes active and overwhelms the helical solution, producing an inter-orbital spin-triplet even-parity state with an $s$-wave gap structure (in band space), see Fig.\;\ref{fig:vafek}(d). As displayed in Fig.\;\ref{fig:vafek}(c), spin-fluctuations and Hund's pairing agree well in this case, because the onsite direct attraction dominates over the weak momentum-dependent parts of Eq.\;(\ref{eq:Veff}).
Further details and parameter dependence are discussed in the SM\;\cite{Supplementary}
Figure\;\ref{fig:vafek}(e) shows the importance of nonzero $\lambda_{so}$ within Hund's pairing, yielding vanishing eigenvalue for $\lambda_{so}\rightarrow 0$. Finally, Fig.\;\ref{fig:vafek}(f) displays the $T$ dependence of the pairing channels in orbital- and spin-space from Hund's pairing corresponding to the components given in Table\;\ref{tab:VafChu_selfcons_OP_channels}. The dominant $A_{1g}$ channel $\frac{1}{2} ([\Delta]^{xz\downarrow}_{yz\uparrow}+[\Delta]^{xz\uparrow}_{yz\downarrow})$ can be written as 
\begin{equation}
    (U'-J) \sum_{\boldsymbol{k}'} \big(\expval{c_{-\kv'yz\down}c_{\kv'xz\up}} + \expval{c_{-\kv'yz\up}c_{\kv'xz\down}} \big),
    \label{eq:vafek_A1_gtriplet}
\end{equation}
highlighting its orbital-singlet, spin-triplet structure,
in agreement with earlier works\;\cite{Vafek17,Cheung_2019}. In summary, non-nested multiorbital band structures generally exhibit agreement between the pairing strengths and gap structures obtained by Hund's pairing and spin-fluctuation pairing for $J/U>\frac{1}{3}$. However, for $J/U<\frac{1}{3}$ only the latter method enables superconductivity. 
\begin{table}[tb]
\caption{Relevant combinations of order parameter components for the two-orbital model shown in Fig.\;\ref{fig:vafek} transforming as irreducible representations (IRs) of the point group $D_{4h}$, see also SM\;\cite{Supplementary}. Orbital- and spin-structure are indicated by their singlet (S) or triplet (T) character.}
\begin{center}
\begin{tabular}{l|l|c|c|c}
\hline \hline
& Channel & Orbital & Spin & IR \\ \hline 
\begin{minipage}{0.4cm} \vspace{0.05cm} \textcolor[RGB]{47,125,85}{\Huge$\bullet$} \end{minipage}& $\frac{1}{2} ([\Delta]^{xz\downarrow}_{yz\uparrow} + [\Delta]^{xz\uparrow}_{yz\downarrow})$ & S &  T & $A_{1g}$ \\ 
\begin{minipage}{0.4cm} \vspace{0.05cm} \textcolor[RGB]{210,105,30}{\Huge$\bullet$} \end{minipage}& $\frac{1}{2} ([\Delta]^{xz\uparrow}_{xz\downarrow} + [\Delta]^{yz\uparrow}_{yz\downarrow} )$ &  T & S & $A_{1g}$\\\hline \hline
\end{tabular}
\end{center}
\label{tab:VafChu_selfcons_OP_channels}
\end{table}

\subsection{Nested band structure}

\begin{figure*}[tb]
    \centering
    \includegraphics[width=\linewidth]{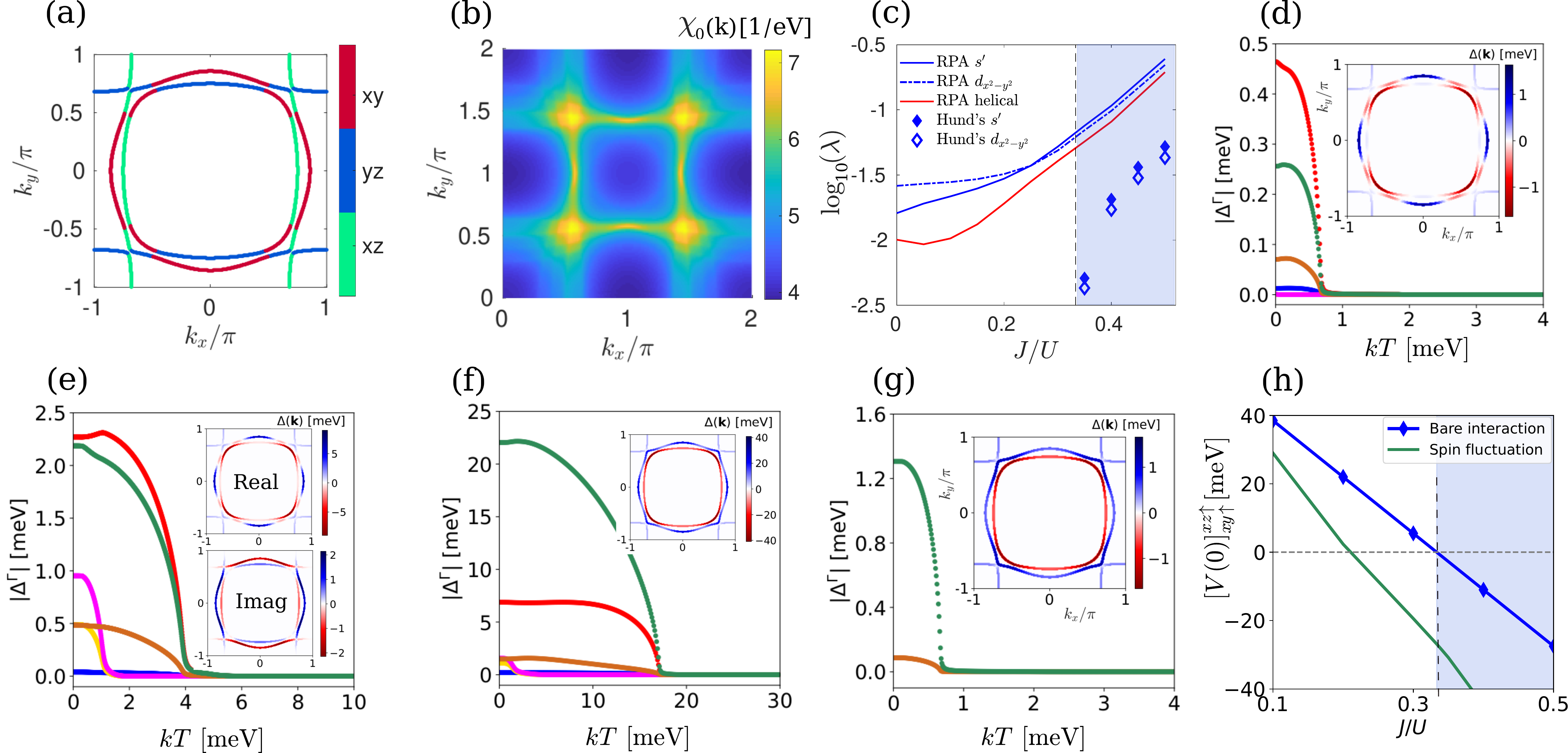}
    \caption{(a) Fermi surface of the three-orbital model of Eq.\;(\ref{eq:H0}). Majority orbital content indicated by the color code. (b) Momentum dependence of the bare static susceptibility $\chi_0^{zz}(\qv)$. (c) Eigenvalues of leading and subleading solutions to the LGE, as a function of $J/U$ for $U=100\;\mathrm{meV}$. Spin-fluctuation (Hund's) mediated pairing is shown by lines (symbols). Shaded blue region indicates regime of attractive Hund's pairing, $J/U>\frac{1}{3}$. (d-f) Orbital- and spin-structure components of the pairing gap versus $T$ as obtained from spin-fluctuation pairing ($U=110\;\mathrm{meV}$) for $J/U=0.3$ (d), $J/U=0.35$ (e), and $J/U=0.4$ (f). See Table\;\ref{tab:selfcons_OP_channels} for the detailed definition and color code of each channel. The insets in (d-g) display the gap structures on the Fermi surface. In (e) the insets represent real (top, $s'$) and  imaginary (bottom, $d_{x^2-y^2}$) parts of the leading $s'+id$ gap function at $T=0$. (g) Orbital- and spin-structure of the gap as obtained from Hund's pairing with $J/U=0.5$. Note much smaller gap scale (and $T_c$) compared to (f), despite larger $J/U$ in (g). (h) Onsite triplet-pairing component $\left[V(0)\right]_{xy\uparrow}^{xz\uparrow}$ [driving the green curves in (d-g)] vs. $J/U$ for spin-fluctuations (green line) and Hund's pairing (blue line), revealing an enhancement of onsite attractions by spin fluctuations. }
    \label{fig:SRO}
\end{figure*}

Next, we turn to a different band representative of cases that do exhibit some nesting. We are not concerned with rare perfectly-nested bands, but rather with bands exhibiting a degree of approximate finite-momentum nesting as typically occurs in many unconventional superconductors. As a concrete, timely example, we apply a band relevant for \sruo\;\cite{Cobo16,RomerPRL}. The normal state Hamiltonian is
\begin{eqnarray}
 H_0(\kv)&=&\left( \begin{array}{ccc}
  \xi_{xz}(\kv) & -i\sigma\lambda_{so}/2 & i\lambda_{so}/2 \\
  i\sigma\lambda_{so}/2 & \xi_{yz}(\kv) & -\sigma\lambda_{so}/2\\
  -i\lambda_{so}/2&-\sigma\lambda_{so}/2 &  \xi_{xy}(\kv)
  \end{array}\right),
  \label{eq:H0}
\end{eqnarray}
in the pseudospin basis $[c_{\kv,xz,\sigma},c_{\kv,yz,\sigma},c_{\kv,xy,\overline\sigma}]$. The dispersion relations are $\xi_{xz/yz}(\kv)=-2t_{1/2}\cos k_x-2t_{2/1}\cos k_y-\mu $
and $\xi_{xy}(\kv)=-2t_3(\cos k_x+\cos k_y)-4t_4\cos k_x \cos k_y-2t_5(\cos 2k_x+\cos 2k_y)-\mu
$ with $(t_1,t_2,t_3,t_4,t_5,\mu)=(88,9,80,40,5,109)\;\mathrm{meV}$ and atomic SOC $\lambda_{so} {\bf L} \cdot {\bf S}\mathrm{,}\;\lambda_{so}=20\;\mathrm{meV}$. Band parameters were chosen in ranges appropriate for \sruo\cite{Cobo16,RomerPRL}. The Fermi surface is shown in Fig.\;\ref{fig:SRO}(a), and the associated spin susceptibility with well-defined nesting peaks is displayed in Fig.\;\ref{fig:SRO}(b). As evident from the blue lines in Fig.\;\ref{fig:SRO}(c), spin-fluctuation pairing generates leading even-parity nodal $s'\;\;(A_{1g})$ or $d_{x^2-y^2}\;\;(B_{1g})$ gap structures throughout the entire $J/U$ range. By $s'$ we refer to the fact that the gap structure is nodal $s$-wave. A subleading odd-parity helical solution (red line) is also depicted in Fig.\;\ref{fig:SRO}(c). In the large-$J$ region, Hund's pairing with significantly smaller eigenvalues sets in, as seen by the diamond symbols in Fig.\;\ref{fig:SRO}(c). The critical temperatures in this regime scale as $T_c \propto e^{-1/\lambda}$ showing a  dominance of spin-fluctuation mediated pairing, originating from important contributions from the fluctuation terms in the pairing kernel, Eq.\;(\ref{eq:Veff}).
More details and examination of parameter dependence are given in the SM\;\cite{Supplementary}

To obtain a detailed understanding of the differences between the gap structures from the two mechanisms,
we turn to selfconsistent solutions of the full gap equation in orbital- and spin-space decomposed into the appropriate irreducible representations (IRs)\;\cite{Ramires2019,Senechal_2019,Huang19}. See Table\;\ref{tab:selfcons_OP_channels} for the relevant components, and SM for the full list\;\cite{Supplementary}. Focusing on the results from spin-fluctuation pairing, Fig.\;\ref{fig:SRO}(d-f) show the $T$-dependence of the non-zero orbital channels at the large ratios $J/U=0.3\mathrm{,}\;J/U=0.35\mathrm{,}\;\mathrm{and}\;J/U=0.4$\;\footnote{Note that for the particular case of \sruo, this amounts to assuming unphysically large Hund's exchange: Constrained RPA calculations find $J/U\simeq\;0.1-0.2$\;\cite{Mravlje11,Vaugier12,Miyake_cRPA}; we nevertheless evaluate the model for large $J$ in order to discuss Hund's pairing states.}. Figure\;\ref{fig:SRO}(d) reveals a dominant orbital-triplet, spin-singlet $A_{1g}$ structure of the nodal $s'$ solution (red curve) generated from spin-fluctuations at $J/U=0.3$\;\cite{RomerPRL}. Inside the (mean-field) Hund's regime, i.e. for $J/U=0.35$, the orbital-singlet spin-triplet (green curve) channel is substantial, but has not yet surpassed the orbital-triplet spin-singlet channel (red curve). As seen from the insets in Fig.\;\ref{fig:SRO}(e), the final $T=0$ gap structure is of the $s'+id_{x^2-y^2}$ form. This composite time-reversal symmetry broken state is expected from the LGE solutions (Fig.\;\ref{fig:SRO}(c)) revealing that the $s'$ and $d_{x^2-y^2}$ channels are nearly degenerate at this $J/U$. At even larger $J/U$, as seen from Fig.\;\ref{fig:SRO}(f), spin-fluctuations favor the orbital-singlet, spin-triplet gap structure and the gap in band space becomes nodeless with sign-changes between the different Fermi sheets. 
\begin{table}[b]
\caption{Same as Table\;\ref{tab:VafChu_selfcons_OP_channels} for the three-orbital model Eq.\;(\ref{eq:H0}). We include the form factors (FF) for each channel, obtained from the projection of the gap into the different basis functions with $f_\pm(\boldsymbol{k})=\cos{2k_x}\pm\cos{2k_y}$.}
\begin{center}
\begin{tabular}{c|l|c|c|c|c}
\hline \hline
& Channel & FF & Or & Sp & IR \\ \hline 
\begin{minipage}{0.4cm} \vspace{0.05cm} \textcolor[RGB]{255, 0, 0}{\Huge$\bullet$} \end{minipage} & $\frac{1}{2\sqrt{3}} ([\Delta]^{xz\uparrow}_{xz\downarrow} + [\Delta]^{yz\uparrow}_{yz\downarrow} -2[\Delta]^{xy\uparrow}_{xy\downarrow})$ & $f_+(\boldsymbol{k})$ & T &  S &  $A_{1g}$ \\
\begin{minipage}{0.4cm} \vspace{0.05cm} \textcolor[RGB]{0, 0, 255}{\Huge$\bullet$} \end{minipage} &$\frac{1}{4} \big(i[\Delta]^{xz\downarrow}_{xy\downarrow} -i [\Delta]^{xz\uparrow}_{xy\uparrow} - [\Delta]^{yz\uparrow}_{xy\uparrow} - [\Delta]^{yz\downarrow}_{xy\downarrow} \big)$ & $f_+(\boldsymbol{k})$ & S & T & $A_{1g}$ \\
\begin{minipage}{0.4cm} \vspace{0.05cm} \textcolor[RGB]{255, 215,0}{\Huge$\bullet$} \end{minipage}&$\frac{1}{3} ([\Delta]^{xz\uparrow}_{xz\downarrow} + [\Delta]^{yz\uparrow}_{yz\downarrow} + [\Delta]^{xy\uparrow}_{xy\downarrow})$ & $f_-(\boldsymbol{k})$ & T & S & $B_{1g}$ \\
\begin{minipage}{0.4cm} \vspace{0.05cm} \textcolor[RGB]{255, 0, 255}{\Huge$\bullet$} \end{minipage}&$\frac{1}{4} \big(i[\Delta]^{xz\downarrow}_{xy\downarrow} -i[\Delta]^{xz\uparrow}_{xy\uparrow}  + [\Delta]^{yz\uparrow}_{xy\uparrow} + [\Delta]^{yz\downarrow}_{xy\downarrow} \big)$ & $1$ & S & T & $B_{1g}$ \\
\begin{minipage}{0.4cm} \vspace{0.05cm} \textcolor[RGB]{47,125,85}{\Huge$\bullet$} \end{minipage}&$\frac{1}{4} \big( i [\Delta]^{xz\downarrow}_{xy\downarrow}- i [\Delta]^{xz\uparrow}_{xy\uparrow} - [\Delta]^{yz\uparrow}_{xy\uparrow} - [\Delta]^{yz\downarrow}_{xy\downarrow} \big)$ & $1$ & S &  T & $A_{1g}$ \\
\begin{minipage}{0.4cm} \vspace{0.05cm} \textcolor[RGB]{210,105,30}{\Huge$\bullet$} \end{minipage}&$\frac{1}{3} ([\Delta]^{xz\uparrow}_{xz\downarrow} + [\Delta]^{yz\uparrow}_{yz\downarrow} + [\Delta]^{xy\uparrow}_{xy\downarrow})$ & $1$ & T & S & $A_{1g}$\\
 \hline \hline
\end{tabular}
\end{center}
\label{tab:selfcons_OP_channels}
\end{table}
The solution of the gap equation within Hund's pairing is substantially simpler: the orbital structure valid throughout the regime $\frac{1}{3}<J/U<\frac{1}{2}$ is shown in Fig.\;\ref{fig:SRO}(g) with the associated momentum gap structure displayed in the inset. In the Hund's mechanism, the orbital-singlet, spin-triplet $A_{1g}$ channel involving inter-orbital pairing between $xy$- and $xz/yz$-orbitals  dominates the pairing, as seen from Fig.\;\ref{fig:SRO}(g). By comparing Figs.\;\ref{fig:SRO}(f) and \ref{fig:SRO}(g) it is evident that the gap structures generated by the two distinct mechanisms become similar when $J/U\simeq\;\frac{1}{2}$. Our selfconsistent solutions confirm, however, that even in this regime of $U$ and $J$, the $T_c$'s of the Hund's pairing mechanism are substantially lower than those from spin-fluctuations, in agreement with Fig.\;\ref{fig:SRO}(c)\;\footnote{For larger values of $\lambda_{so}$, the order parameters in orbital- and spin-space become qualitatively similar at lower values of $J/U$ when comparing Hund's pairing vs. spin-fluctuation pairing, but mean-field Hund's pairing still produce significantly lower $T_c$ values.}.

The extent of the nominal Hund's pairing regime can be renormalized by higher-order scattering processes\;\cite{Hoshino_2015}. We demonstrate this explicitly in Fig.\;\ref{fig:SRO}(h) displaying the $J/U$-dependence of the dominant onsite pairing channel for 1) mean-field Hund's pairing given by $\frac{U}{2}(1-\frac{3J}{U})$, and 2) the {\it onsite part} of the spin-fluctuation pairing, $V(\rv=0)$. For the latter case, to 2nd order in interactions the pairing between  $xz$- and $xy$-orbitals for same-spin electrons is proportional to
\begin{eqnarray}
    V^{(2)}(\rv=0)\simeq
    (U'-J)&&-UU'\sum_\qv [ \chi_0^{xz}(\qv) +\chi_0^{xy}(\qv)] \nonumber \\
     -(U')^2\sum_\qv\chi_0^{yz}(\qv)&&-(U'-J)^2\sum_\qv\chi_0^{yz}(\qv),
    \label{eq:v2}
\end{eqnarray}
  where we have included for simplicity only the dominant intra-orbital susceptibilities, see SM\;\cite{Supplementary}. This demonstrates that spin fluctuations can induce onsite attraction in the regime $J/U<\frac{1}{3}$, where the bare interaction is repulsive.  Importantly, however, this onsite channel is still substantially weaker than (non-local) channels driven by spin-fluctuation finite-momentum pair scattering processes, as evident from Fig.\;\ref{fig:SRO}(c,d,e)\;\footnote{For ferromagnetic fluctuations the renormalization of the onsite channel should more reliably capture changes to the pairing vertex from spin-fluctuations.}.

\section{Discussion and conclusions}

Inter-orbital pair states are candidates for the superconducting ground state of many materials\;\cite{Spalek_2001,Han_2004,Sakai_2004,Kubo_2007,Lee_2008,Xia_2008,Puetter_2012,Zegrodnik_2014,Hoshino_2015,Vafek17,Cheung_2019,suh2019,Boker_2019,clepkens_PRB_2021}. We have discussed the stability of such states within two different,  widely used methods for superconducting pairing. For non-nested band structures with finite SOC, onsite Hund's pairing at the mean-field level agrees well with spin-fluctuation mediated pairing in the large-$J$ region. By contrast, for correlated materials with bands exhibiting some degree of finite-momentum nesting, pronounced susceptibility contributions remain important, from small $J$ up to and including inside much of the the large-$J$ Hund's regime, and can lead to qualitatively different gap structures as seen e.g. from comparing Figs.\;\ref{fig:SRO}(e) and \ref{fig:SRO}(g). This is relevant even though fluctuations enhance the Hund's regime for purely local pairing. We have discussed the latter findings for a band structure relevant for \sruo, but the results are expected to remain valid for generic bands with some degree of finite-momentum nesting including e.g. iron-based superconductors. Our calculations therefore serve as a cautionary note against the indiscriminate application of the Hund's pairing approach, and suggest that in most unconventional superconductors of current interest, orbital-singlet, spin-triplet states of this type are not realized.

\begin{acknowledgments}
This work is supported by Novo Nordisk Foundation grant NNF20OC0060019 (M.~R.). A.~T.~R. and B.~M.~A. acknowledge support from the Independent Research Fund Denmark, grant number 8021-00047B. P.~J.~H. was supported by the U.S. Department of Energy under Grant No. DE-FG02-05ER46236. 
\end{acknowledgments}

%
\ifarXiv
    

\section*{Supplementary material}

# Supplementary Material for Superconductivity in multiorbital systems with repulsive interactions: Hund's vs. spin fluctuation pairing

Mercè Roig,[1] Astrid T. Rømer,[1] Andreas Kreisel,[2] P. J. Hirschfeld,[3] and Brian M. Andersen[1]

[1]*Niels Bohr Institute, University of Copenhagen, 2100 Copenhagen, Denmark*
[2]*Institut für Theoretische Physik, Universität Leipzig, D-04103 Leipzig, Germany*
[3]*Department of Physics, University of Florida, Gainesville, Florida 32611, USA*

In this supplementary material, we provide details on the classification of the pairing channels in orbital- ans spin-space (Sec. S1), the calculations using the linearized gap equation (Sec. S2), the pairing kernel within spin-fluctuation mediated pairing and the derivation of a renormalized effective interaction (Sec. S3), and additional parameter dependence of the results for both the two- (Sec. S4) and the three-orbital models (Sec. S5) discussed in the main text.

## S1. SELFCONSISTENT SOLUTION IN ORBITAL- AND SPIN-SPACE

In this section of the supplementary material, we discuss the formalism used for the full selfconsistent solution of the gap equation in orbital and spin space. The mean field gaps are labelled by two orbital and spin indices and we define the index $\tilde{\mu}_i = (\mu_i, s_i)$ combining orbital $(\mu)$ and spin $(s)$ labels

$$[\Delta_{\mathbf{k}}]^{\tilde{\mu}_i}_{\tilde{\mu}_j} = \sum_{\mathbf{k}', \tilde{\mu}_{\mathbf{k}}, \tilde{\mu}_{\mathbf{l}}} [V(\mathbf{k}, \mathbf{k}')]^{\tilde{\mu}_i, \tilde{\mu}_l}_{\tilde{\mu}_j, \tilde{\mu}_k} \langle c_{-\mathbf{k}' \tilde{\mu}_l} c_{\mathbf{k}' \tilde{\mu}_k} \rangle. \tag{S1}$$

Introducing the Fourier transform of the multi-orbital interaction we can write the gap as a function of the real space pairing and the basis functions for the different neighbors in the point group $D_{4h}$, including up to 28 neighbors and on-site interactions in the spin-fluctuation mediated pairing case. Consequently, we can separate the odd and the even irreducible representations depending on the symmetry of the form factor $g^{\Gamma}_{\mathbf{k}}$,

$$[\Delta_{\mathbf{k}}]^{\tilde{\mu}_i}_{\tilde{\mu}_j} = \sum_{\Gamma \in \mathrm{IR}} g^{\Gamma}_{\mathbf{k}} [\Delta_{\Gamma}]^{\tilde{\mu}_i}_{\tilde{\mu}_j} = \sum_{\Gamma_g \in \mathrm{IR}} g^{\Gamma_g}_{\mathbf{k}} [\Delta_{\Gamma_g}]^{\tilde{\mu}_i}_{\tilde{\mu}_j} + \sum_{\Gamma_u \in \mathrm{IR}} g^{\Gamma_u}_{\mathbf{k}} [\Delta_{\Gamma_u}]^{\tilde{\mu}_i}_{\tilde{\mu}_j}, \tag{S2}$$

where $\Gamma_g$ and $\Gamma_u$ correspond to the even and the odd basis functions, respectively, and

$$[\Delta_{\Gamma}]^{\tilde{\mu}_i}_{\tilde{\mu}_j} = \sum_{\mathbf{k}', \tilde{\mu}_k, \tilde{\mu}_l} [V_{\mathbf{k}', \Gamma}]^{\tilde{\mu}_i, \tilde{\mu}_l}_{\tilde{\mu}_j, \tilde{\mu}_k} \langle c_{-\mathbf{k}' \tilde{\mu}_l} c_{\mathbf{k}' \tilde{\mu}_k} \rangle. \tag{S3}$$

The possible mean field gap structures are classified by the irreducible representations in orbital and spin space [1]. Each of these structures can then be combined with a irreducible representation of $D_{4h}$ classifying the momentum structure.

Since the order parameter fulfills $[\Delta_{\mathbf{k}}]^{\tilde{\mu}_i}_{\tilde{\mu}_j} = -[\Delta_{-\mathbf{k}}]^{\tilde{\mu}_j}_{\tilde{\mu}_i}$, the matrices coupling to even form factors satisfy $[\Delta_{\Gamma_g}]^{\tilde{\mu}_i}_{\tilde{\mu}_j} = -[\Delta_{\Gamma_g}]^{\tilde{\mu}_j}_{\tilde{\mu}_i}$. Using the notation $\hat{\Delta} = i(\hat{\psi} + \hat{\mathbf{d}} \cdot \boldsymbol{\sigma}) \sigma_y$ for the order parameter, we can separate the spin-singlet and spin-triplet combinations.

In the three-orbital model, the generators for the orbital structure are given by the Gell-Mann matrices, which we define as

$$\lambda_1 = \begin{pmatrix} 0 & 1 & 0 \\ 1 & 0 & 0 \\ 0 & 0 & 0 \end{pmatrix}, \qquad \lambda_2 = \begin{pmatrix} 0 & -i & 0 \\ i & 0 & 0 \\ 0 & 0 & 0 \end{pmatrix}, \qquad \lambda_3 = \begin{pmatrix} 1 & 0 & 0 \\ 0 & -1 & 0 \\ 0 & 0 & 0 \end{pmatrix},$$

$$\lambda_4 = \begin{pmatrix} 0 & 0 & 1 \\ 0 & 0 & 0 \\ 1 & 0 & 0 \end{pmatrix}, \qquad \lambda_5 = \begin{pmatrix} 0 & 0 & -i \\ 0 & 0 & 0 \\ i & 0 & 0 \end{pmatrix}, \qquad \lambda_6 = \begin{pmatrix} 0 & 0 & 0 \\ 0 & 0 & 1 \\ 0 & 1 & 0 \end{pmatrix},$$

$$\lambda_7 = \begin{pmatrix} 0 & 0 & 0 \\ 0 & 0 & -i \\ 0 & i & 0 \end{pmatrix}, \qquad \lambda_8 = \frac{1}{\sqrt{3}} \begin{pmatrix} 1 & 0 & 0 \\ 0 & 1 & 0 \\ 0 & 0 & -2 \end{pmatrix}, \qquad \mathbb{1}_{\lambda} = \begin{pmatrix} 1 & 0 & 0 \\ 0 & 1 & 0 \\ 0 & 0 & 1 \end{pmatrix}. \tag{S4}$$

We derive how the orbital ($\mathbb{1}_\lambda$, $\boldsymbol{\lambda}$) and the spin generators ($\mathbb{1}_\sigma$, $\boldsymbol{\sigma}$) transform under each point group element of $D_{4h}$, where $\boldsymbol{\lambda}$ refers to the Gell-Mann matrices and $\boldsymbol{\sigma}$ to the Pauli matrices:

$$\sigma_x = \begin{pmatrix} 0 & 1 \\ 1 & 0 \end{pmatrix}, \qquad \sigma_y = \begin{pmatrix} 0 & -i \\ i & 0 \end{pmatrix}, \qquad \sigma_z = \begin{pmatrix} 1 & 0 \\ 0 & -1 \end{pmatrix}, \qquad \mathbb{1}_\sigma = \begin{pmatrix} 1 & 0 \\ 0 & 1 \end{pmatrix}. \tag{S5}$$

The irreducible representation corresponding to each mean field channel is given by the product of the orbital and the spin representation and the outcome of this procedure is shown in Table S1. All gap structures are classified according to the orbital triplet or singlet structure and whether it corresponds to spin-triplet or spin-singlet gap. The overall spin- and orbital structure is classified by the irreducible representation (IR) of $D_{4h}$, see last column of Table I.
In a few cases the total representation is reducible (see row 10, 12, 13 and 15 of Table I). In these cases, taking particular combinations of the reducible representations, an irreducible representation of $D_{4h}$ can be obtained, as shown in Table S2.) We show only the classification scheme of the matrices coupling to even parity gap structures, since odd-parity solutions are not stabilized in the selfconsistent procedure of the models under consideration in this work.

In the case of the two-orbital model also belonging to the $D_{4h}$ group, we follow a similar procedure, see Table S3. Likewise, we find a number of possible orbital-singlet, spin triplet solutions as well as orbital-triplet, spin singlet solutions. Also in this case, we restrict the discussion to even parity solutions. The obtained structures of Table S3 are in agreement with Ref. 2.

Table S1. Combinations of the order parameter matrix components for the three-orbital model transforming as irreducible representations of the point group $D_{4h}$ that couple with an even form factor $g_{-\boldsymbol{k}}^{\Gamma_g} = g_{\boldsymbol{k}}^{\Gamma_g}$. We include how each combination couples to orbital and spin space, and specify if they correspond to triplet or singlet.

| $\hat{\Delta}_{\Gamma_g}$ | Orbital Space $\lambda$ | | Spin space $\sigma$ | | IR |
|---|---|---|---|---|---|
| $\frac{1}{3}([\Delta_{\Gamma_g}]_{xz\downarrow}^{xz\uparrow} + [\Delta_{\Gamma_g}]_{yz\downarrow}^{yz\uparrow} + [\Delta_{\Gamma_g}]_{xy\downarrow}^{xy\uparrow})$ | Triplet, Intra | $\mathbb{1}_\lambda$ | Singlet | $\mathbb{1}_\sigma$ | $A_{1g}$ |
| $\frac{1}{2}([\Delta_{\Gamma_g}]_{xz\downarrow}^{xz\uparrow} - [\Delta_{\Gamma_g}]_{yz\downarrow}^{xz\downarrow})$ | Triplet | $\lambda_1$ | Singlet | $\mathbb{1}_\sigma$ | $B_{2g}$ |
| $\frac{1}{2}([\Delta_{\Gamma_g}]_{xz\downarrow}^{xz\uparrow} - [\Delta_{\Gamma_g}]_{yz\downarrow}^{yz\uparrow})$ | Triplet, Intra | $\lambda_3$ | Singlet | $\mathbb{1}_\sigma$ | $B_{1g}$ |
| $\frac{1}{2}([\Delta_{\Gamma_g}]_{xy\downarrow}^{xz\uparrow} - [\Delta_{\Gamma_g}]_{xy\uparrow}^{xz\downarrow})$ | Triplet | $\lambda_4$ | Singlet | $\mathbb{1}_\sigma$ | $E_{yg}(i)$ |
| $\frac{1}{2}([\Delta_{\Gamma_g}]_{xz\downarrow}^{yz\uparrow} - [\Delta_{\Gamma_g}]_{xy\uparrow}^{yz\downarrow})$ | Triplet | $\lambda_6$ | Singlet | $\mathbb{1}_\sigma$ | $E_{xg}(i)$ |
| $\frac{1}{2\sqrt{3}}([\Delta_{\Gamma_g}]_{xz\downarrow}^{xz\uparrow} + [\Delta_{\Gamma_g}]_{yz\downarrow}^{yz\uparrow} - 2[\Delta_{\Gamma_g}]_{xy\downarrow}^{xy\uparrow})$ | Triplet, Intra | $\lambda_8$ | Singlet | $\mathbb{1}_\sigma$ | $A_{1g}$ |
| $\frac{1}{2}([\Delta_{\Gamma_g}]_{yz\uparrow}^{xz\uparrow} + [\Delta_{\Gamma_g}]_{yz\downarrow}^{xz\downarrow})$ | Singlet | $\lambda_2$ | Triplet | $\sigma_y$ | $E_{yg}(ii)$ |
| $\frac{i}{2}([\Delta_{\Gamma_g}]_{yz\downarrow}^{xz\uparrow} + [\Delta_{\Gamma_g}]_{yz\uparrow}^{xz\downarrow})$ | Singlet | $\lambda_2$ | Triplet | $\sigma_z$ | $A_{1g}$ |
| $\frac{-i}{2}([\Delta_{\Gamma_g}]_{yz\uparrow}^{xz\uparrow} - [\Delta_{\Gamma_g}]_{yz\downarrow}^{xz\downarrow})$ | Singlet | $\lambda_2$ | Triplet | $\sigma_x$ | $E_{xg}(ii)$ |
| $\frac{1}{2}([\Delta_{\Gamma_g}]_{xy\uparrow}^{xz\uparrow} + [\Delta_{\Gamma_g}]_{xy\downarrow}^{xz\downarrow})$ | Singlet | $\lambda_5$ | Triplet | $\sigma_y$ | Reducible |
| $\frac{i}{2}([\Delta_{\Gamma_g}]_{xy\downarrow}^{xz\uparrow} + [\Delta_{\Gamma_g}]_{xy\uparrow}^{xz\downarrow})$ | Singlet | $\lambda_5$ | Triplet | $\sigma_z$ | $E_{yg}(iii)$ |
| $\frac{-i}{2}([\Delta_{\Gamma_g}]_{xy\uparrow}^{xz\uparrow} - [\Delta_{\Gamma_g}]_{xy\downarrow}^{xz\downarrow})$ | Singlet | $\lambda_5$ | Triplet | $\sigma_x$ | Reducible |
| $\frac{1}{2}([\Delta_{\Gamma_g}]_{xy\uparrow}^{yz\uparrow} + [\Delta_{\Gamma_g}]_{xy\downarrow}^{yz\downarrow})$ | Singlet | $\lambda_7$ | Triplet | $\sigma_y$ | Reducible |
| $\frac{i}{2}([\Delta_{\Gamma_g}]_{xy\downarrow}^{yz\uparrow} + [\Delta_{\Gamma_g}]_{xy\uparrow}^{yz\downarrow})$ | Singlet | $\lambda_7$ | Triplet | $\sigma_z$ | $E_{xg}(iii)$ |
| $\frac{-i}{2}([\Delta_{\Gamma_g}]_{xy\uparrow}^{yz\uparrow} - [\Delta_{\Gamma_g}]_{xy\downarrow}^{yz\downarrow})$ | Singlet | $\lambda_7$ | Triplet | $\sigma_x$ | Reducible |

Table S2. Irreducible representations coupling to an even form factor obtained by combining the four reducible representations in Table S1.

| $\hat{\Delta}_{\Gamma_g}$ | Combined spin and orbital | Orbital | Spin | IR |
|---|---|---|---|---|
| $\frac{1}{4}([\Delta_{\Gamma_g}]_{xy\uparrow}^{xz\uparrow} + [\Delta_{\Gamma_g}]_{xy\downarrow}^{xz\downarrow} - i([\Delta_{\Gamma_g}]_{xy\uparrow}^{yz\uparrow} - [\Delta_{\Gamma_g}]_{xy\downarrow}^{yz\downarrow}))$ | $\lambda_5\sigma_y + \lambda_7\sigma_x$ | Singlet | Triplet | $A_{2g}$ |
| $\frac{1}{4}([\Delta_{\Gamma_g}]_{xy\uparrow}^{xz\uparrow} + [\Delta_{\Gamma_g}]_{xy\downarrow}^{xz\downarrow} + i([\Delta_{\Gamma_g}]_{xy\uparrow}^{yz\uparrow} - [\Delta_{\Gamma_g}]_{xy\downarrow}^{yz\downarrow}))$ | $\lambda_5\sigma_y - \lambda_7\sigma_x$ | Singlet | Triplet | $B_{2g}$ |
| $\frac{1}{4}(-i([\Delta_{\Gamma_g}]_{xy\uparrow}^{xz\uparrow} - [\Delta_{\Gamma_g}]_{xy\downarrow}^{xz\downarrow}) + [\Delta_{\Gamma_g}]_{xy\uparrow}^{yz\uparrow} + [\Delta_{\Gamma_g}]_{xy\downarrow}^{yz\downarrow})$ | $\lambda_5\sigma_x + \lambda_7\sigma_y$ | Singlet | Triplet | $B_{1g}$ |
| $\frac{1}{4}(-i([\Delta_{\Gamma_g}]_{xy\uparrow}^{xz\uparrow} - [\Delta_{\Gamma_g}]_{xy\downarrow}^{xz\downarrow}) - [\Delta_{\Gamma_g}]_{xy\uparrow}^{yz\uparrow} - [\Delta_{\Gamma_g}]_{xy\downarrow}^{yz\downarrow})$ | $\lambda_5\sigma_x - \lambda_7\sigma_y$ | Singlet | Triplet | $A_{1g}$ |

Table S3. Combinations of the order parameter matrix components for the two-orbital model transforming as irreducible representations of the point group $D_{4h}$ that couple with an even form factor $g_{-\mathbf{k}}^{\Gamma_g} = g_{\mathbf{k}}^{\Gamma_g}$. We include how each combination couples to orbital and spin space, and specify if they correspond to triplet or singlet.

| $\Delta_{\Gamma_g}$ | Orbital Space $\tau$ | Spin space $\sigma$ | IR |
|---|---|---|---|
| $\frac{1}{2}([\Delta_{\Gamma_g}]_{xz\uparrow}^{xz\uparrow} + [\Delta_{\Gamma_g}]_{yz\downarrow}^{yz\uparrow})$ | Triplet, Intra $\mathbb{1}_\tau$ | Singlet $\mathbb{1}_\sigma$ | $A_{1g}$ |
| $\frac{1}{2}([\Delta_{\Gamma_g}]_{xz\uparrow}^{xz\uparrow} - [\Delta_{\Gamma_g}]_{yz\downarrow}^{yz\uparrow})$ | Triplet $\tau_1$ | Singlet $\mathbb{1}_\sigma$ | $B_{1g}$ |
| $\frac{1}{2}([\Delta_{\Gamma_g}]_{yz\uparrow}^{xz\uparrow} - [\Delta_{\Gamma_g}]_{yz\downarrow}^{xz\uparrow})$ | Triplet, Intra $\tau_3$ | Singlet $\mathbb{1}_\sigma$ | $B_{2g}$ |
| $\frac{1}{2}([\Delta_{\Gamma_g}]_{yz\uparrow}^{xz\uparrow} + [\Delta_{\Gamma_g}]_{yz\downarrow}^{xz\downarrow})$ | Singlet $\tau_2$ | Triplet $\sigma_y$ | $E_{yg}$ |
| $\frac{1}{2}([\Delta_{\Gamma_g}]_{yz\uparrow}^{xz\uparrow} - [\Delta_{\Gamma_g}]_{yz\downarrow}^{xz\downarrow})$ | Singlet $\tau_2$ | Triplet $\sigma_z$ | $E_{xg}$ |
| $\frac{1}{2}([\Delta_{\Gamma_g}]_{yz\downarrow}^{xz\downarrow} + [\Delta_{\Gamma_g}]_{yz\uparrow}^{xz\uparrow})$ | Singlet $\tau_2$ | Triplet $\sigma_x$ | $A_{1g}$ |

## S2. LINEARIZED GAP EQUATION

We also analyze the Hund's pairing and spin-fluctuation pairing using the linearized gap equation (LGE)

$$-\frac{1}{(2\pi)^2} \int_{FS} d\mathbf{k}'_f \frac{1}{v(\mathbf{k}'_f)} \Gamma_{l,l'}(\mathbf{k}_f, \mathbf{k}'_f) \Delta_{l'}(\mathbf{k}'_f) = \lambda \Delta_l(\mathbf{k}_f), \tag{S6}$$

where $\Gamma_{l,l'}(\mathbf{k}_f, \mathbf{k}'_f)$ is the pairing kernel, Eq. (S10), projected to band- and spin-space [3–5] and $\mathbf{k}_f$ denotes a wave vector on the Fermi surface. The Fermi speed is given by $v(\mathbf{k}_f)$. For Hund's pairing, only the first (constant) interaction term is present, while for spin-fluctuation pairing, also the momentum-dependent terms in Eq.. (S10) arising from exchange of spin-fluctuations are considered. Within this method, the leading superconducting instability is given by the gap function $\Delta_l(\mathbf{k}_f)$ with the largest eigenvalue $\lambda$. The (pseudo)spin classification is encoded in the subscripts $l, l' = 0, x, y, z$ which refer to the components of the $\mathbf{d}(\mathbf{k})$-vector [4, 6].

## S3. SPIN-FLUCTUATION MEDIATED PAIRING

In order to derive an expression for the multi-orbital pairing mediated by spin-fluctuations in the general case where spin-orbit coupling is important, we define a generalized susceptibility given by:

$$[\chi_0]_{\mu_3 s_3, \mu_4 s_4}^{\mu_1 s_1, \mu_2 s_2}(\mathbf{q}, i\omega_n) = \frac{1}{N} \int_0^\beta d\tau e^{i\omega_n \tau} \sum_{\mathbf{k},\mathbf{k}'} \langle T_\tau c_{\mathbf{k}-\mathbf{q}\mu_1 s_1}^\dagger(\tau) c_{\mathbf{k}\mu_2 s_2}(\tau) c_{\mathbf{k}'+\mathbf{q}\mu_3 s_3}^\dagger(0) c_{\mathbf{k}'\mu_4 s_4}(0) \rangle_0. \tag{S7}$$

dependent on four independent orbital ($\mu$) and spin ($s$) indices. In the normal state, this becomes

$$[\chi_0]_{\mu_3 s_3, \mu_4 s_4}^{\mu_1 s_1, \mu_2 s_2}(\mathbf{q}, i\omega_n) = -\frac{1}{N} \sum_{\mathbf{k}} \sum_{n_1, n_2} [M_{n_1, n_2}(\mathbf{k}, \mathbf{q})]_{\mu_3 \sigma_3, \mu_4 \sigma_4}^{\mu_1 \sigma_1, \mu_2 \sigma_2} \frac{f(\xi_{\mathbf{k}-\mathbf{q}, n_1, \sigma_1}) - f(\xi_{\mathbf{k}, n_2, \sigma_2})}{i\omega_n + \xi_{\mathbf{k}-\mathbf{q}, n_1, \sigma_1} - \xi_{\mathbf{k}, n_2, \sigma_2}}, \tag{S8}$$

with

$$[M_{n_1, n_2}(\mathbf{k}, \mathbf{q})]_{\mu_3 \sigma_3, \mu_4 \sigma_4}^{\mu_1 \sigma_1, \mu_2 \sigma_2} = [u_{n_1 \sigma_1}^{\mu_1 s_1}(\mathbf{k} - \mathbf{q})]^* [u_{n_2 \sigma_3}^{\mu_3 s_3}(\mathbf{k})]^* u_{n_2 \sigma_2}^{\mu_2 s_2}(\mathbf{k}) u_{n_1 \sigma_4}^{\mu_4 s_4}(\mathbf{k} - \mathbf{q}), \tag{S9}$$

where $u_{n\sigma}^{\mu s}(\mathbf{k})$ is the Eigenvector of the transformation from orbital and electronic spin basis ($\mu, s$) to band and pseudo-spin basis ($n, \sigma$).

Summing up all ladder and bubble diagrams to infinite order in the bare interactions $U, U', J, J'$, we arrive at the pairing interaction

$$\left[V(\mathbf{k}, \mathbf{k}')\right]_{\tilde{\mu}_j, \tilde{\mu}_k}^{\tilde{\mu}_i, \tilde{\mu}_l} = \left[U\right]_{\tilde{\mu}_j, \tilde{\mu}_k}^{\tilde{\mu}_i, \tilde{\mu}_l} + \left[U \frac{1}{1 - \chi_0 U} U\right]_{\tilde{\mu}_j, \tilde{\mu}_l}^{\tilde{\mu}_i, \tilde{\mu}_l}(\mathbf{k} + \mathbf{k}') - \left[U \frac{1}{1 - \chi_0 U} U\right]_{\tilde{\mu}_j, \tilde{\mu}_k}^{\tilde{\mu}_i, \tilde{\mu}_k}(\mathbf{k} - \mathbf{k}'), \tag{S10}$$

which is stated as Eq. (3) in the main text. In this expression $[\chi_0]$ and $[U]$ denote matrices of dimension $36 \times 36$ encoding all possible orbital and spin possibilities at every order in the diagrammatic expression. The bare interaction matrices are given by

$$\left[U\right]_{\mu\overline{\sigma}, \mu\sigma}^{\mu\sigma, \mu\overline{\sigma}} = U, \qquad \left[U\right]_{\mu\overline{\sigma}, \nu\sigma}^{\nu\sigma, \mu\overline{\sigma}} = U', \qquad \left[U\right]_{\nu\overline{\sigma}, \nu\sigma}^{\mu\sigma, \nu\overline{\sigma}} = J', \qquad \left[U\right]_{\nu\sigma, \nu\sigma}^{\mu\sigma, \mu\sigma} = J, \qquad \left[U\right]_{\nu\sigma, \mu\sigma}^{\mu\sigma, \nu\sigma} = U' - J, \tag{S11}$$

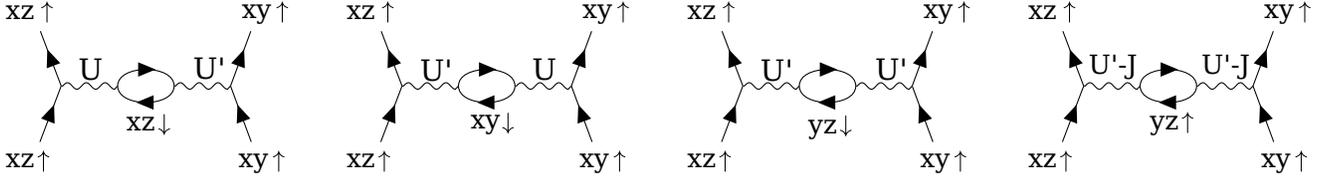

Figure S1. Second order diagrams relevant for interactions between same-spin electrons on orbital $xz$ and $yz$ when restricting to intra-orbital susceptibilities only.

where $\mu$ and $\nu$ denote different orbitals. All symmetry-related elements are included explicitly in the construction of the bare interaction matrix $[U]^{\tilde{\mu}_1,\tilde{\mu}_2}_{\tilde{\mu}_3,\tilde{\mu}_4}$.

If we restrict to diagrams where only the intra-orbital susceptibility components are taken into account, i.e. restricting to $\chi_0^\mu(\mathbf{q}) = [\chi_0]^{\mu s,\mu s}_{\mu s,\mu s}(\mathbf{q})$, where all orbitals are identical, we get the second order correction to the onsite interaction:

$$\tilde{V}^{(2)}(\mathbf{r} = 0) = -UU'\sum_{\mathbf{q}}[\chi_0^{xz}(\mathbf{q}) + \chi_0^{xy}(\mathbf{q})] - (U')^2\sum_{\mathbf{q}}\chi_0^{yz}(\mathbf{q}) - (U' - J)^2\sum_{\mathbf{q}}\chi_0^{yz}(\mathbf{q}). \tag{S12}$$

The four relevant diagrams are shown in Fig. S1. Since the intra-orbital susceptibilities are positive numbers and $U, U' > 0$, Eq. (S12) clearly shows how a uniform susceptibility provides a renormalization of the onsite interaction, enabling effective onsite attraction also when the bare interaction $U' - J$ is repulsive.

In the specific case of $\lambda_{so} = 20$ meV treated in Fig. 2 of the main text, we have $\sum_{\mathbf{q}}\chi_0^{xz}(\mathbf{q}) = \sum_{\mathbf{q}}\chi_0^{yz}(\mathbf{q}) = 2.84$ eV$^{-1}$ and $\sum_{\mathbf{q}}\chi_0^{xy}(\mathbf{q}) = 4.24$ eV$^{-1}$.

To third order, the correction from bubble diagrams reads

$$\begin{aligned}\tilde{V}^{(3)}(\mathbf{r} = 0) = &U^2(U' - J)[\chi_{0,a}^2 + \chi_{0,b}^2 + \chi_{0,a}\chi_{0,b}]\\ &+ (U')^2(U' - J)[3\chi_{0,a}^2 + 5\chi_{0,a}\chi_{0,b}]\\ &+ (U' - J)^3[\chi_{0,a}^2 + 2\chi_{0,a}\chi_{0,b}]\\ &+ UU'(U' - J)[7\chi_{0,a}^2 + 3\chi_{0,a}\chi_{0,b}]\end{aligned} \tag{S13}$$

where we introduced the abbreviated notation

$$\chi_{0,a}^2 = \sum_{\mathbf{q}}(\chi_0^{xz}(\mathbf{q}))^2 = \sum_{\mathbf{q}}(\chi_0^{yz}(\mathbf{q}))^2 \tag{S14}$$

$$\chi_{0,b}^2 = \sum_{\mathbf{q}}(\chi_0^{xy}(\mathbf{q}))^2 \tag{S15}$$

$$\chi_{0,a}\chi_{0,b} = \sum_{\mathbf{q}}\chi_0^{xz}(\mathbf{q})\chi_0^{xy}(\mathbf{q}) = \sum_{\mathbf{q}}\chi_0^{yz}(\mathbf{q})\chi_0^{xy}(\mathbf{q}) \tag{S16}$$

In the specific case of $\lambda_{so} = 20$ meV treated in Fig. 2 of the main text, we have $\chi_{0,a}^2 = 8.69$ eV$^{-2}$, $\chi_{0,b}^2 = 18.20$ eV$^{-2}$ and $\chi_{0,a}\chi_{0,b} = 12.15$ eV$^{-2}$. From the third order expression, Eq. (S16), we observe that the third order diagrams counteracts the effect of the second order diagrams stated in Eq. (S12), adding a positive contribution to the onsite effective interaction. We note that the full interaction vertex entering the gap equation, Eq. (1), must be symmetrized to yield a definite parity for orbital exchange. The fully renormalized onsite interaction displayed by the green curve in Fig. 2 (h) of the main paper is symmetrized according to

$$\left[V(\mathbf{r} = 0)\right]^{xz\uparrow}_{xy\uparrow} = \left[U\right]^{xz\uparrow,xy\uparrow}_{xy\uparrow,xz\uparrow} + \left[V(\mathbf{r} = 0)\right]^{xz\uparrow,xy\uparrow}_{xy\uparrow,xz\uparrow} - \left[V(\mathbf{r} = 0)\right]^{xz\uparrow,xz\uparrow}_{xy\uparrow,xy\uparrow}. \tag{S17}$$

## S4.  NON-NESTED CASE

For the non-nested band structure defined by the Hamiltonian

$$H_0(\mathbf{k}) = \begin{pmatrix} \mu - a\mathbf{k}^2 + bk_xk_y & c(k_x^2 - k_y^2) - i\sigma\lambda_{so} \\ c(k_x^2 - k_y^2) + i\sigma\lambda_{so} & \mu - a\mathbf{k}^2 - bk_xk_y \end{pmatrix} \tag{S18}$$

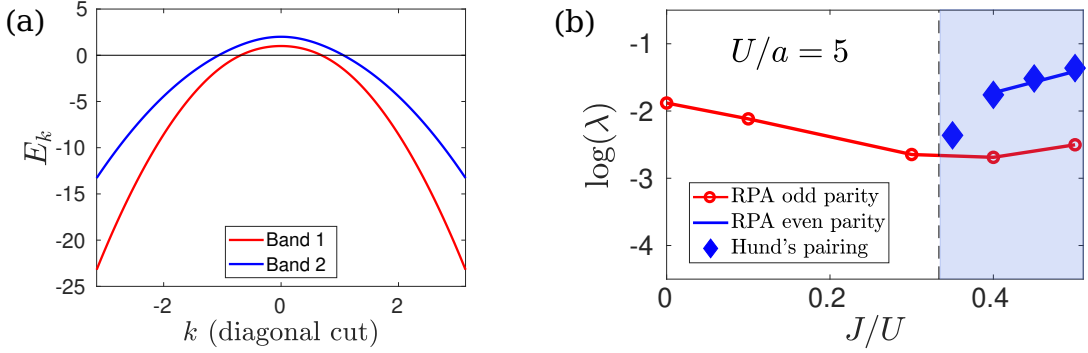

Figure S2. (a) Diagonal cut through the two energy bands of the Hamiltonian stated in Eq. (S18) for parameters: $a = 1$, $b = c = \lambda_{so} = 0.5$, $\mu = 1.5$. (b) Leading eigenvalue of the LGE as a function of $J/U$ for $U/a = 5$. Results from spin-fluctuation pairing (Hund's pairing) are indicated by solid lines (diamonds). Light blue region indicates regime of attractive onsite Hund's pairing, $J/U > \frac{1}{3}$.

the energy bands is shown in Fig. S2(a). The top of the hole pockets is of the order of $E/a = 1$ away from the Fermi level. In the main paper, we treat the case of $U/a = 1$ in Figs. 1 (c,d), displaying solutions to the linearized gap equation (LGE). Due to the weak susceptibility, the Hund's regime $J/U > \frac{1}{3}$ is dominated by the superconducting gap solutions originating from onsite attractions, i.e. the first term of the pairing interaction stated in Eq. (S10), as shown in the main paper Fig. 1(c) in the case $U/a = 1$ and this remains to be the case for $U/a = 5$, as shown in Fig. S2(b). For this model, the critical value of Hubbard $U$ is roughly $U_c/a \simeq 20$, depending on the size of the Hund's coupling. In the regime very close to the critical interaction strengths, the superconducting solutions from spin-fluctuation mediated pairing, i.e. the odd-parity helical solutions, will dominate over the $A_{1g}$ solutions of the Hund's pairing, even inside the regime of large $J/U > \frac{1}{3}$.

## S5. NESTED CASE: DEPENDENCE ON THE COUPLING STRENGTHS

For the nested case, we investigate the dependence of the bare interaction strength by repeating the calculation of leading instabilities of the linearized gap equation for smaller Coulomb interaction $U$. In Fig. S3 (a-c), we show the evolution of the leading eigenvalues for spin-fluctuation mediated pairing as well as Hund's pairing upon increasing values of $J/U$ in the cases $U = 50, 80, 100$ meV. The latter case corresponds to Fig. 2(c) of the main paper. Note that for the smallest value of $U = 50$ meV (Fig. S3(a)), the eigenvalue of the spin-fluctuation mediated pairing approaches the eigenvalues of the Hund's pairing for $J/U > \frac{1}{3}$. This is due to the fact that the nesting structures of $\chi_0(\mathbf{q})$ are suppressed in the pairing kernel at $\mathcal{O}(U^2)$, see Eq. (S10).

For all three cases of $U$, we plot the leading instability at $J/U = 0.2, 0.3, 0.35$ in Fig. S3(d-f). In the case of small $U$ (Fig. S3(a,d)), the $d_{x^2-y^2}$ is leading for $J/U < \frac{1}{3}$, whereas for $J/U = 0.35$, the $s$-wave is leading, albeit a strong gap suppression on the middle pocket ($\gamma$-pocket), reminiscent of the nodal structure arising in spin-fluctuation mediated pairing. As $U$ is increased, Fig. S3(e,f), the $s$-wave solution moves to smaller values of $J/U = 0.3$ ($< \frac{1}{3}$), due to the renormalized onsite interaction as discussed in the main paper Fig. 2 (h). At the same time, nesting features become increasingly important for stronger values of $U$, leading to enhanced gap suppressions at the $\gamma$-pocket also in the Hund's regime. To highlight the structure of the RPA spin susceptibility in all three cases, we show in Fig. S3(g-j) the nesting peaks, and how these increase upon increasing values of $U$ as well as $J/U$.

Finally, in Fig. S4, we show how the magnetic instability is approached as a function of increasing $U$. For larger values of $J/U$, the critical $U$ is lowered, as seen from the four cases plotted in Fig. S4.

---

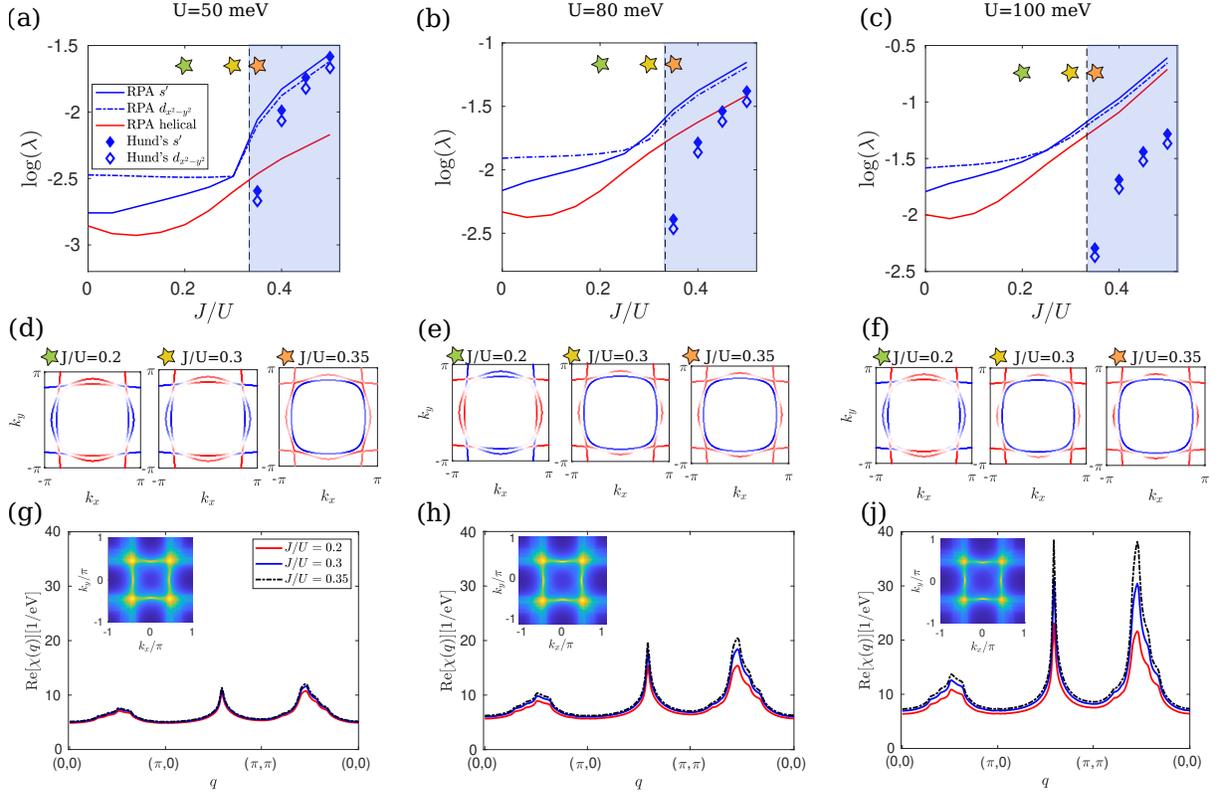

Figure S3. (a-c) Eigenvalues of leading and subleading solutions to the LGE Eq. (5) of the main paper, as a function of $J/U$ for $U = 50$ meV (a), $U = 80$ meV (b), and $U = 100$ meV (c). Spin-fluctuation (Hund's) mediated pairing is shown by lines (symbols). Shaded blue region indicates regime of attractive Hund's pairing, $J/U > \frac{1}{3}$. (d-f) Leading solutions of the RPA calculation for three different values of $J/U = 0.2, 0.3, 0.35$ as indicated by colored stars in the case of $U = 50$ meV (d), $U = 80$ meV (e), and $U = 100$ meV (f). (g-i) Longitudinal spin susceptibility for three different values of $J/U = 0.2, 0.3, 0.35$ in the case of $U = 50$ meV (g), $U = 80$ meV (h), and $U = 100$ meV (i) plotted along the path $(0,0) - (\pi,0) - (\pi,\pi) - (0,0)$. The insets show the longitudinal susceptibility at $J/U = 0.35$ in the respective cases.

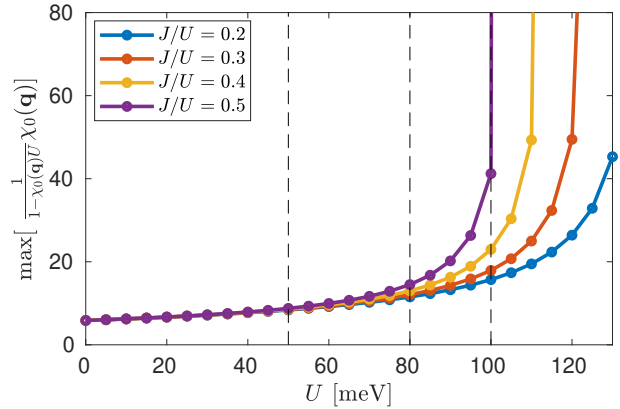

Figure S4. Maximum value of the generalized susceptibility as a function of $U$ for four different values of $J/U = 0.2, 0.3, 0.4, 0.5$. The dashed lines mark the $U$-values applied in Fig. S3.


\fi


\end{document}